# The Effect of Education on Smoking Decisions in the United States

Sang Truong

DePauw University

sangtruong_2021@depauw.edu

December 7, 2018

**Abstract:** This paper explores the link between education and the decision to start smoking and the decision to quit smoking. Data is gathered from IPUMS CPS and the Centers for Disease Control and Prevention. Probit analysis (with the use of probability weight and robust standard error) indicates that every additional year of education will reduce the 2.3 percentage point of the smoking probability and will add 3.53 percentage point in quitting likelihood, holding home restriction, public restriction, cigarette price, family income, age, gender, race, and ethnicity constant. I believe that the tobacco epidemic is a serious global issue that may be mitigated by using careful regulations on smoking restriction and education.

**Acknowledgments:** I want to express my gratitude to Professor Humberto Barreto for his time and effort to support me. I am also thankful for having Benton Turner, Gustavo Kasmanas, and other fellows to support me with Stata and to give me constructive comments on my paper.

**I. Introduction**

Smoking can generate a tremendous negative effect on the smoker and his surroundings. Smoking has a detrimental consequence on health. Indeed, according to the American Lung Association, there are about 600 chemicals in a cigarette. When they are burned, they are transformed into 7000 different chemicals and about 70 of them cause cancer. Smoking damages almost every organ in the body and causes several diseases in the lung and heart in addition to countless cancers. Smoking decreases life expectancy by 10 years on average. Besides, smokers hurt their surroundings through secondhand smoking. Every year, secondhand smoking causes 41 000 deaths through heart diseases and lung cancer (Centers for Diseases Control and Prevention, 2018).

Even with the knowledge about the negative effects of smoking, many smokers fall into the trap of thinking that it is too late to quit because their body is damaged already. However, according to the American Lung Association, there is evidence suggesting that quit smoking is beneficial at every age. After about 2 weeks of quit smoking, the risk of having a heart attack of the smoker starts to drop and his lung function begins to improve. After about 1 year after quitting, "the added risk of coronary heart disease is half that of a smoker's". After 15 years after quitting, "the risk of coronary heart disease [of former smoker] is the same as that of a nonsmoker" (American Lung Association, 2018). Regardless of how long one has smoked, it is never too late to quit smoking.

However, when an individual commit to smoke, it is very hard for him to give up smoking due to the addictive chemicals in the cigarette. For example, nicotine, a chemical in the cigarette can produce temporary pleasure through physical and mood-altering effects in the brain of the smoker. These pleasures make the smoker want to smoke more and lead to nicotine dependence. Once the smoker depends on nicotine, stop smoking causes withdrawal symptoms, including anxiety and irritability, that prevent him from quitting smoking, even when he notices the significant degrade in his health condition.

Nowadays, the global battle against smoking and its consequences does not only belong to health practitioner or scientists but everyone, including the government and the whole academia society. In this war, there are two broad but important questions: (1) What should we do to prevent smoking behavior? And (2) What should we do to promote quitting behavior? To answer these two questions, we need to understand the motivations behind the decision to start smoking and the decision to quit smoking. There are several different factors that determine these decisions. Among factors that can be regulated, education is one of the most important one that affects smoking and quitting decision. Therefore, the scope of this paper is to explore the relationship between education year and smoking behavior. My research question is ***"How strong education can affect smoking and quitting decision?"*** I find that, on average, an individual with a higher number of education year has a lower probability of smoking and a higher probability of quitting. The protective effect of education is significantly enforced by the restriction of smoking at home. This result may provide some suggestions for public policy to reduce tobacco consumption. In the war against tobacco addiction, education may be a better weapon than we thought.

The rest of this paper includes 4 parts. Section II presents the Theory of Rational Addiction by Becker and Murphy and several articles about the relationship between education and smoking/quitting probability. In addition, section II also provides the information from previous studies about the effect of statistical controls (such as race, gender, and cigarette price) on education and/or smoking/quitting likelihood as a rationale for my decision of selecting these statistical controls. Section III presents (A) the necessary information about data acquired from IPUMS CPS as well as Centers for Disease Control and Prevention and (B) two probit models, their marginal analyses, and the result interpretation. Section IV includes the conclusion about the economic impact of the result as well as the limitation of this study from a policy-making point of view.

In this paper, "smoke" and "quit" are used as perfect substitutions for "starting to smoke for the first time" and "quit smoking," respectively.

**II. Literature review**

The decision of smoking, which cause detrimental effects on health condition, seems to be irrational. However, many scholars believe that the smoking decision of an individual is an attempt to optimize his utility regarding his current condition. Becker and Murphy (1994) establish the Theory of Rational Addiction to address the case of the smoking decision. According to this theory, the effect of smoking on smoker's utility can be decomposed into three main components: current benefits, future losses, and adjustment costs:

- Current benefits are utilities that are derived in the time smoking action occur, including (1) the sensational pleasure that comes from the taste or smell of cigarette or (2) the satisfaction from the output of smoking (such as increasing the concentration, being accepted by friends, or rebelling against parents).
- Future losses are losses of utility in the future due to the damaging effect of smoking. Future losses are discounted to the present value so that it can be compared with the current benefits. The current value of future losses that a person uses to compare to the current benefits depends on (1) his ability to fully recognize the magnitude of his future losses and (2) his life expectancy.
- Adjustment costs are the losses of utility when the smoker decides to reduce or quit smoking. These costs present the withdrawal symptoms that came from the pharmacological effect of nicotine on smoker's brain.

The total effect of current benefits, future losses, and adjustment costs will determine the smoking behavior of an individual. Only if total effect is positive, he will decide to smoke.

Many studies indicate that education has a strong correlation with smoking and quitting behavior. When an individual does not fully realize the huge future losses of smoking activity, he will likely to wrongly decide to smoke. The ability to understand the future consequence of smoking depends heavily on education. Indeed, Gilman (2008) and Paraje (2017) concludes that higher education associates with a lower probability of smoking and a higher probability of quitting smoking. Carla (2009) conducts a survey to understand the relationship between the major study and the smoking probability in the population of 6492 students in the University of Minnesota. The result shows that the field of study is associated with the smoking probability. Indeed, with the control of demographic, psychosocial, and other health behavior variable, the highest smoking rate are among those who are communication, language, or culture study (37.4%) and the lowest rate are among mathematics, engineering, and science major (21.0%). Carla et al suggest that a better understanding about future losses generated by the smoking activity of science students significantly decreases their smoking likelihood. These literatures provide a consistent result showing that education is an important predictor for smoking behavior.

Since age, gender, race, ethnicity, and income correlate with education, it is necessary to include them in the models as statistical controls to avoid omitted variable bias. In addition to the correlation to education, these variables also have their own effect on smoking/quitting behavior. I elaborate some information about the relationship between age, gender, race, ethnicity, and income here to rationalize my decision to put them into my model:

- **Age:** As one gets older, he will less tolerate with the future losses, ceteris paribus, since the discounted factor gets closer to 1. Therefore, smoking behavior can correlate to age. In addition, as one gets older, his ability to derive pleasure from current benefits of smoking activity will likely to decrease. Jarvis (2004) argues that smoking behavior usually occurs at the early teenage years because smoking is a way to take risk and to rebel against parents. The teenager smokers usually use smoking as a symbolic act to convey that he is now mature. As one gets older, the demand to convey this message decrease. Therefore, age is an important predictor for smoking behavior.

- **Gender:** Teenagers and adolescents are not only the ones who use smoking as a symbolic act. Woman can also use smoking to convey feminist messages. Historically, men are more likely to smoke than women. The difference in smoking behavior by gender deeply rest on gender inequality. Until the 19th century, smoking is the symbol of masculine. During the Victorian period, women start to smoke as a symbolic action for independence (Eliot, 2001). Advertisements of tobacco had not targeted women until the evolution of feminism. Those advertisements that target woman promote smoking as "a symbol of emancipation", and "a torch of freedom" (Hunt, 2004). Nowadays, there is still a gender gap, and tobacco company continues to use it to promote their products.

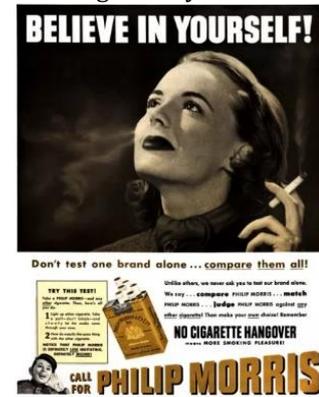

**Figure 1.** A cigarette advertisement targeting woman. Smoking is used to convey feminism.

- **Income:** In addition to education, age, and gender, income can affect smoking behavior. Indeed, higher income leads to a higher purchasing power, associating with the higher possibility of tobacco consumption. With higher income, the rich have more freedom to explore a more varied consumption possibility, including tobacco. It intuitively makes sense that a person who does not need to worry about the next meal will more likely to purchase cigarette than the one who worries about her monthly electric bill, ceteris paribus. Perelman (2017) performs a logistic regression on adolescents in 6 different cities in European to understand the likelihood of being an experimenter, a weekly smoker, and a daily smoker. After statistically controlling for personal traits (such as citizenship, gender, age, parental smoking status) and family social-economic status (family influence, subjective social-economic status, and parental education), result indicates that "adolescents in the highest income quintile were more likely to be smoking experimenters, weekly smokers, and daily smokers than those in the lowest quintile. They also consumed more cigarettes per month."

- **Race and Ethnicity:** Race can affect smoking behavior. The inherent cultural differences that come from different races are shown to shape the smoking and quitting decision. Messer (2011) shows that "compared with non-Hispanic Whites, smaller proportions of African Americans, Asian Americans/Pacific Islanders, and Hispanics/Latinos had ever smoked" after holding age, gender, education, and poverty level constant. In addition, this study also showed that "significantly fewer African Americans reported long-term quitting." Although there is not a straightforward and intuitive explanation for the difference in smoking motivation across race, race could possibly correlate to other factors, such as the probability of having smoking restriction at home, that determine the smoking and quitting motivation. Indeed, Mai (2018) shows that there is a correlation between the smoking regulation at home and race with statically control for parent smoking status. Using data from IPUMS CPS during the period of 2010 – 2011 and 2014 – 2015, Mai shows that Asian parents are most likely to set the smoking restriction at home (94%), while multiracial parents are least likely to do so (77%).

Due to the complicated relationship between education, race, and smoking restriction as studied by Mai (2018), it is necessary to include smoking restriction in the regression model to avoid confounding. However, another reason to include smoking restriction in the regression model is its important role to determine smoking/quitting behavior. Indeed, smoking rules and restrictions in home and in public areas can reduce the smoking probability. The smoking bans will make smoking practice become inconvenient. Therefore, when one lives in a society in which smoking is unacceptable or morally wrong, he will have a lower likelihood to smoke. Indeed, Chapman (1999), Farkas (1999), Gilpin (1999), and Wakefield (2000) show that restrictions on smoking at home and in public places reduce levels of smoking in adults and teenagers. In addition, the bands in public have a stronger effect on home bans. Onder (2012) shows that smoking probability increases with the decrease of school regulation. In addition, Wakefield (200) shows that school bans have very little effect on smoking probability unless they are strongly enforced. However, since smoking behavior occurs during the teenager and adolescence period, in which individual is affected by their friends more than by their parents or professors (Harris, 1998), the effect of these regulations can vary significantly depending on other factors, such as the relationship between individual and his parents. Regardless of the effectiveness, there is a consistent result from the above articles showing that some forms of smoking regulation reduce the smoking likelihood.

Tobacco price has a strong effect on cigarette consumption due to the law of demand and supply. Onder (2012) uses 2 logit models to regress the probability of smoking on the natural log of price on 15 957 students from 202 schools in Turkey. With statistical controls such as demographic information, family characteristic (education of parents, the smoking status of parents), the exposure level to cigarette, and school regulation, Onder shows that the price of cigarette has a strong, negative effect on the smoking probability: 10% increase in cigarette price decreases the smoking probability by 11.69%. Moreover, when tax is used instead of price, the result is similar: If the government increases the tobacco tax by 10%, the smoking probability of the young is expected to fall by 9%. The price sensitivity of female is higher than that of male. This result implies that the tobacco price may impact the smoking behavior.

In summary, education has a strong correlation with smoking/quitting behavior. However, it also correlates with other factors, such as age, gender, race, ethnicity, and income. Through carefully review previous studies, I realize that race and ethnicity correlate with smoking restriction, which is known as an important factor that affects the smoking/quitting decision. Therefore, I decide to add smoking restriction as another additional statistical control for education. Last but not least, tobacco price is added into the regression model in an attempt to improve the performance of the model.

**III. Empirical Strategy**

This paper focuses on the effect of education on (1) the decision of smoking and (2) the decision of quitting smoking. Two theoretical models that are employed to study the effect of education on the probability of smoking and the likelihood of quitting:

- **Smoking model:** P(*smoke*) = f (*resthome, restpub, lnprice, educyear, famincome, age, agesq, male, white, black, asian, hisp*)
- **Quitting model:** P(*quit*) = f (*resthome, restpub, lnprice, educyear, famincome, age, male, white, black, asian, hisp*)

**1. Dependent Variables**

a. Dependent variable for smoking model

The decision of smoking is a dummy variable: at the time the data were collected, the responder can only have 2 answers:

- He has not decided to start smoking, meaning he is a "never smoker."
- He has already made the decision to start smoking before. He could be a "current smoker" (either "daily smoker" or "non-daily smoker") or a "former smoker."

The variable chosen from CPS IPUMS to represent the smoking decision is Smoking Record (TSMKER). According to IPUMS CPS, TSMKER is a recoded that classifies respondents as "daily smoker", "non-daily smoker", "former smoker", or a "never smoker". The fact that TSMKER is a recoded indicating that information about smoking record is derived from another variable rather acquiring from the survey. Based on the 4 possible outcomes of TSMKER, a new dummy variable (*smoke*) is constructed to distinguish between "never smoker" and "current smoker".

The information that can best elucidate the smoking motivation is the one collected at the time decision was made. The further from the time decision made, the less accurate the information about personal trails and environment can be used to understand the smoking motivation. For a former smoker, it is very likely that deterministic factors that motivate smoking behavior are already reshaped. Therefore, due to the limitation of the data, I believe that it is too complicated to study the smoking motivation of former smoker. Thus, data about former smokers will not be used to study the smoking motivation. These data about former smokers will be used to study the quitting motivation. As explained above, the "former smoker" is not covered by this dummy variable. More specific:

- If TSMKER = 1 (never smoker), *smoke* = 0;
- If TSMKER = 3 (non-daily smoker) or = 4 (daily smoker), *smoke* = 1; and

b. Dependent variable for quitting decision

Like smoking decision, quitting decision (*quit*) is a dummy variable: Among the responder who had already smoked in the past, there is only 2 possible responds: he does not smoke currently (*quit* = 1) or he is still having the smoking habit (*quit* = 0). The variable chosen from CPS IPUMS to represent the quitting decision is Smoking Record (TSMKER). The data about never smoker will not be used to study quitting motivation.

- If TSMKER = 2 (former smoker), *quit* = 1;
- If TSMKER = 3 or 4 (non-daily and daily smoker), *quit* = 0.

**2. Included independent variables**

**Table 1** describes the abbreviation, explanation, type (dummy or continuous), and the data sources from CPS IPUMS (if applicable) of 12 explanatory variables that are included in the models. **Table 2** presents summary statistics of these variables.

**Table 1.** Information about 12 included independent variables

| X-Variable abbreviation | Explanation | Type | CPS IPUMS sources |
|---|---|---|---|
| resthome | Restrict smoking at home | Dummy, 1 = restrict, 0 = no restrict | TRULEHH |
| restpub | Restrict smoking at the public workplace | Dummy, 1 = restrict, 0 = no restrict | TWKPUBLIC |
| lnprice | Natural logarithm of average cigarette price by states | Continuous variable | From Centers for Diseases Control and Prevention |
| educyear | Number of year of education | Continuous variable | EDUC |
| famincome | Family income of the householder | Continuous variable | FAMINC |
| Age | Age | Continuous variable | AGE |
| Agesq | Age squared | Continuous variable | AGE |
| Male | Male | Dummy, 1 = male, 0 = female | SEX |
| White | Race | Dummy, 1 = white, 0 = not white | RACE |
| Black | Race | Dummy, 1 = black, 0 = not black | RACE |
| Asian | Race | Dummy, 1 = Asian, 0 = not Asian | RACE |
| Hisp | Hispanic | Dummy, 1 = Hispanic, 0 = not Hispanic | HISPAN |

**Table 2.** The summary statistics of 2 dependent variables (smoke and quit) and 12 independent variables.

| Variable | Obs | Mean | Std. Dev. | Min | Max |
|---|---|---|---|---|---|
| smoke | 63,700 | 0.1669 | 0.3729 | 0 | 1 |
| quit | 25,438 | 0.5821 | 0.4932 | 0 | 1 |
| educyear | 102,092 | 13.4698 | 2.6705 | 2.5 | 18 |
| resthome | 55,645 | 0.8586 | 0.3484 | 0 | 1 |
| restpub | 23,848 | 0.8298 | 0.3758 | 0 | 1 |
| lnprice | 102,441 | 1.8257 | 0.1986 | 1.5107 | 2.3185 |
| famincome | 102,441 | 6.5350 | 4.4874 | 0.5 | 15 |
| age | 102,441 | 47.9779 | 17.8749 | 18 | 85 |
| agesq | 102,441 | 2621.3850 | 1794.4450 | 324 | 7225 |
| male | 102,441 | 0.4781 | 0.4995 | 0 | 1 |
| white | 102,441 | 0.8133 | 0.3897 | 0 | 1 |
| black | 102,441 | 0.1038 | 0.3050 | 0 | 1 |
| asian | 102,441 | 0.0514 | 0.2209 | 0 | 1 |
| hisp | 102,441 | 0.1227 | 0.3281 | 0 | 1 |

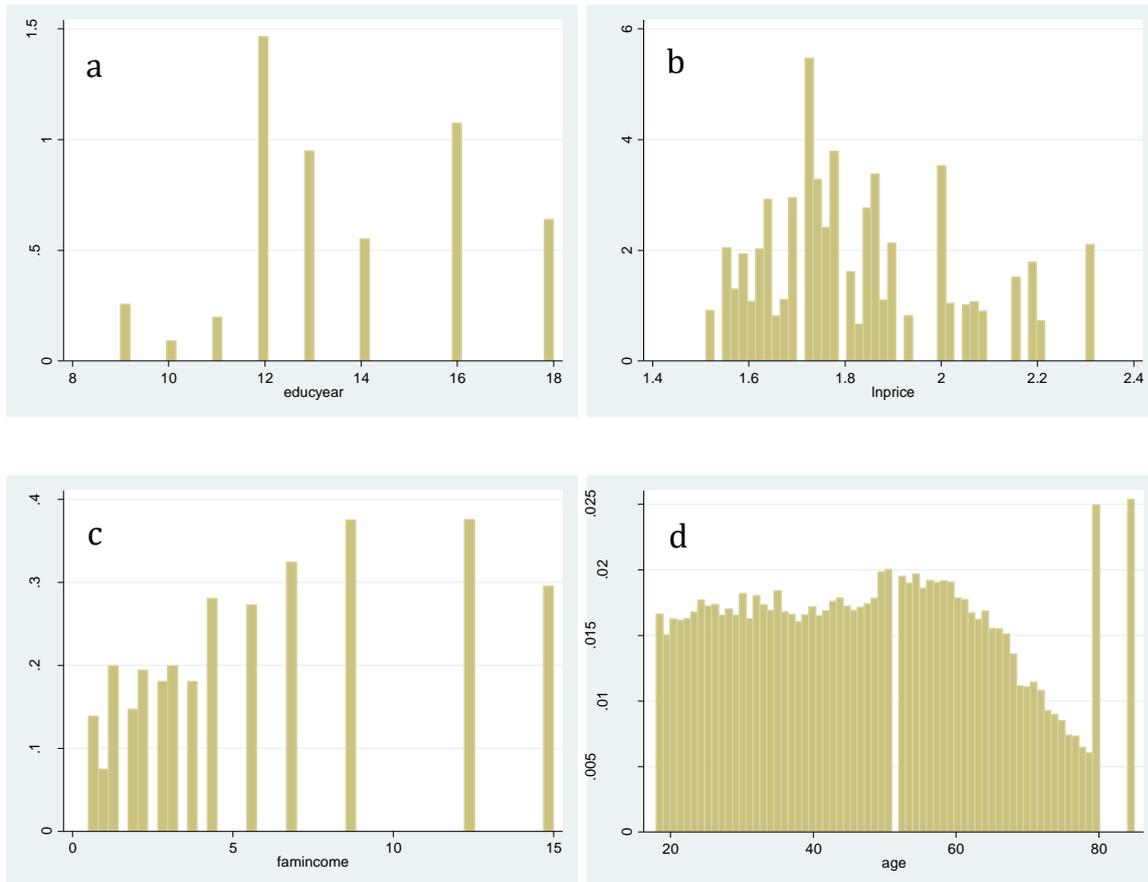

**Figure 2.** Histogram of Education Year (a), Ln price (b), Family Income (c), and Age (d)

*Educyear* (number of year of education) is a continuous variable generating from EDUC from IPUM CPS. Sometimes, the value of EDUC is given as a range (i.e. from grade 1 to grade 4). In these situations, *educyear* is assigned the average value of the range. For example, if EDUC ranges from 1 to 4, *educyear* = 2.5.

Data about average after-tax cigarette price by states in 2015 are acquired from the Centers for Disease Control and Prevention and are incorporated into IPUMS CPS (Centers for Disease Control and Prevention, 2018). The exponential relationship between price and demand motivates the employment of natural log of price (Ln Price) instead of price in the regression models. Indeed, Onder (2012) and Lee (2004) use the natural log of price in their regression models to understand the effect of price on tobacco demand. Since the tobacco demand and smoking/quitting decision are highly correlated, it is necessary to logarithmic transform the cigarette price before putting it into the regression equation. There is no evidence showing the need for arithmetic transformation for other included variables.

The range of *age* studied in this paper is limited to at or above 18 years old due to the fact that 18 is the legal age to smoke in the U.S. As mention before, since smoking behavior has the characteristic of risk talking and rebellion, which is commonly found in the young, teenagers and adolescents have the highest likelihood to smoke. As the *age* increases, the demand for risk and rebellion decrease, leading to a decrease in the smoking decision. Indeed, the smoking probability – age profile (Figure 3) shows a curvature relationship. Therefore, it is necessary to include *agesq* (age squared) in the regression model to faithfully reflect the curvature relationship in the smoking probability – age profile. Since there is no theory or empirical analysis (to the best of my knowledge) indicating that age has a quadratic relationship with quitting decision, *agesq* will not be include in my quitting model in this paper.

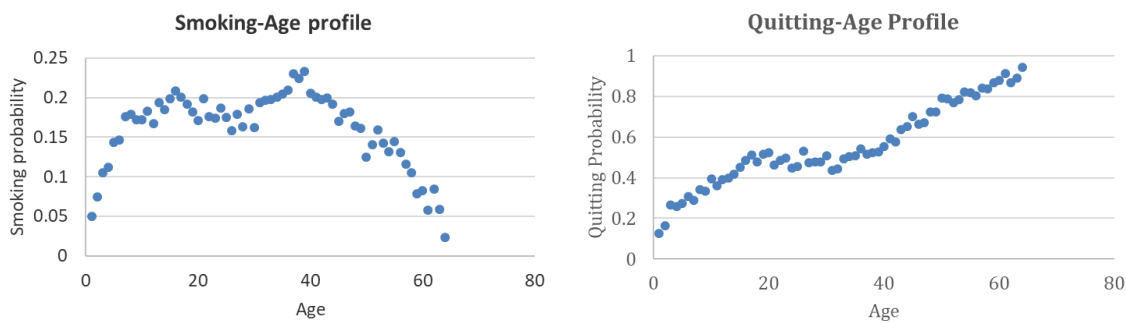

**Figure 3.** Smoking – age and quitting – age profiles

*Resthome*, meaning smoking restriction at home, is a dummy variable that is generated by using TRULEHH variable (smoking rules and restrictions inside your home) from IPUMS CPS. To acquire the information the smoking rules and restriction inside home, the responders are asked the following question: *"Which statement best describes the rules about smoking inside your home? ("home" is where you live; "rules" include any unwritten "rules" and pertain to all people whether or not they reside in the home or are visitors, workmen, etc.)"* There are 3 possible answers:

- TRULEHH = 1: No one is allowed to smoke anywhere inside your home.
- TRULEHH = 2: Smoking is allowed in some places or at sometimes inside your home.
- TRULEHH = 3: Smoking is permitted anywhere inside your home.

I recode TRULEHH to resthome as following:

- If TRULEHH = 1, *resthome* = 1. There is strictly restriction.
- If TRULEHH = 2 or 3, *resthome* = 0. There is no strictly restriction.

Similarly, *restpub*, meaning smoking restriction in the public area, is a dummy variable that is generated based on TWKPUBLIC variable from IPUMS CPS. According to IPUMS CPS, "TWKPUBLIC reports the smoking policy in public areas, such as lobbies, restrooms, and lunch rooms, for the employer of the survey respondent. This variable provides information about whether the employer has a policy and how strong the policy is." To acquire this information, IPUMS CPS ask responders these following questions:

- Is smoking restricted in any way at your place of work? By 'restricted, we mean any limitation on smoking, regardless of who is responsible for that restriction (including the owner, employer, government, union, etc.).
- Which of these best describe the smoking policy at your place of work for indoor public or common areas, such as lobbies, restrooms, and lunch rooms?

There are 4 possible answers:

- TWKPUBLIC = 1: allowed in all public areas
- TWKPUBLIC = 2: allowed in some public areas
- TWKPUBLIC = 3: Not allowed in any public areas
- TWKPUBLIC = 5: No policy

TWKPUBLIC is recoded into *restpub* as following:

- If TWKPUBLIC = 1, 2, or 5, *restpub* = 0.
- If TWKPUBLIC = 3, *restpub* = 1.

*Famincome* (family income) is a continuous variable that is generated based on FAMINC variable from IPUMS CPS. Responders provide the family income information by choosing the income range that best describes his family income. Famincome is recoded from FAMINC by choosing the middle point of the range. For example, if FAMINC ranges from $20 000 to $30 000, *famincome* = $25 000. In addition, before regression analysis, the data about family income is divided by 10 000 to make the estimated coefficient easier to understand.

Lastly, *male* is a dummy variable recoded from SEX variable from IPUMS CPS. *While, black, asian* are 3 dummy variables that are recoded based on RACE variable from IPUMS CPS. Other types of race, such as Asian American, are recoded as based case. These other races only contribute to 4% of the population dataset in this study. *Hisp* is a dummy variable that is generated based on HISPAN to indicate the ethnicity of the responder. HISPAN is recoded to *hisp* as following:

- If HISPAN = 000 (the responder is not Hispanic), *hisp* = 0.
- If HISPAN = 100 (Mexican), 200 (Puerto Rican), 300 (Cuban), 400 (Dominican), 500 (Salvadoran), 411 (Central American excluding Salvadoran), and 412 (South American), *hisp* = 1.

Data for both dependent and independent variables are cleaned by dropping the case of NIU (not in universe), refuse to answer, or do not know the answer. All of the data from IPUMS CPS that are used in this paper are collected at January of 2015. Since the data from IPUMS CPS are acquired through the complex survey, not everyone in the population has an equal likelihood to be chosen. Therefore, the coefficients of every models in this study are estimated with probability weight using WTFINL weight. According to IPUMS CPS, "WTFINL is the final person-level weight that should be used in analyses of basic monthly data." Incorporating final person-level weight into the regression analysis will improve the reliability of the result.

## IV. Results

**Table 3** and **table 4** present the main results of this study. **Table 3** describes smoking models, which is used to understand the importance of factors that affect the smoking decision. **Table 4** describes quitting models, which is used to understand the quitting smoking decision. There are 4 models in each table:

- Model 1 is the simplest model, using probit regression to regress dependent variable (*smoke* or *quit*) on the primary independent variable, *educyear*.
- Model 2 is the longer version of model 1, with the addition of 2 variables: *resthome* and *restpub* (smoking regulation at home and smoking regulation in public, respectively).
- Model 3 is a probit model with the largest set of statistical control. It is the longer model of model 2 and it is the longest probit model in this analysis.
- Model 4 uses OLS method to regress dependent variables on the full set of independent variables.

Since the coefficients of probit regression cannot be interpreted easily, a marginal transformation of the coefficient is needed to understand the meaning of the regression coefficients. For this analysis, the coefficients are transformed into respective average marginal effect. For independent variable $x_i$, its average marginal effect is calculated by taking the average of all marginal effect of $x_i$ at each data point of $x_i$. For example, if $x_i = \{1, 2, 3, 4\}$, then

$$Average\ marginal\ effect\ of\ xi = \frac{\frac{dy}{dxi}(at\ x_i = 1) + \frac{dy}{dxi}(at\ x_i = 2) + \frac{dy}{dxi}(at\ x_i = 3) + \frac{dy}{dxi}(at\ x_i = 4)}{4}$$

The average marginal effect is meaningful for continuous variables $x_i$ because it shows the increase/decrease in the percentage point of smoking/quitting likelihood on average for each additional unit increase in $x_i$. For example, the average marginal effect of education in smoking model 1 with its robust SE (**table 3**) is –0.0213 (0.0005), indicating that every additional year of education reduces the smoking probability by 2.13 percentage point on average, holding nothing else constant. Similarly, the average marginal effect of education in quitting model 1 with its robust SE (**table 4**) is 0.0417 (0.0015), indicating that every additional year of education increases the quitting probability by 4.17 percentage point on average, holding nothing else constant.

It is important to note that since there is only one sample, the coefficients of independent variables reported in **table 3** and **table 4** are just estimators of their true, unbiased value. For example, the coefficient of *educyear* in smoking model 1 and its robust SE is –0.0213 (0.0005), indicating that this coefficient is not static. The robust SE of this coefficient, which is 0.0005, indicates that if I study another sample, I will likely to get another different coefficient for *educyear*. Robust SE can be used to construct a 95% confident interval with 2 units of SE away from the point estimator:

$$-0.0213 +/- 2*0.0005 = (-0.0223; -0.0203)$$

95% of the confident interval constructed in this way will successfully capture the true, unbiased parameter that indicates the true relationship between *educyear* and *smoke* in smoking model 1.

Because there is nothing hold constant in model 1, the above arguments about the relationship between education smoking/quitting decision is vulnerable due to omitted variable bias. Omitted variable bias, or confounding, is the bias in coefficient of the estimator $x_i$ when the variables that are correlated to $x_i$ are omitted from the model. Therefore, it is necessary to add more statistical controls into models 1 to improve the reliability of these models.

The main purpose of introducing model 1, 2, and 3 is to understand how the marginal transformed coefficient of education changes with the addition of statistical controls. The education coefficient of smoking model 1, 2, and 3 with their robust SE are −0.0213 (0.0005), −0. 0223 (0.0010), and −0.0230 (0.0011), respectively. There is almost no economic important change in the magnitude of the education coefficient with the addition of more statistical controls. Indeed, there is only 0.1 percentage point drop in the education coefficient with the introduction of 11 statistical controls in this model. The education coefficient of quitting model 1, 2, and 3 with their robust SE are 0.0417 (0.0015), 0.0406 (0.0029), and 0.0353 (0.0030). Similar to smoking model, there is not economic important change in the coefficient of education of quitting model: There is only 6.4 percentage point change in the education coefficient when 10 statistical controls are added. These economic unimportant changes in the coefficient of education in both smoking and quitting model imply that there is no important relationship between education year and other included statistical controls.

Even though the magnitude of education coefficient does not change a lot, the presence of statistical controls in model 3 improve the reliability of our result by partially protecting it from omitted variable bias. After introducing the statistical controls, I have more confidence to state that, according to smoking model 3, every additional year of education will reduce the 2.3 percentage point of the smoking probability, holding other included independence variable constant. In addition, according to the quitting model, every additional year of education will increase the quitting probability by 3.53 percentage point, holding other included independent variables constants. These results consist with the Theory of Rational Addiction. Indeed, with a higher education year, an individual is likely to better fully recognize the detrimental cost of smoking in the future. Therefore, he will be less tolerant of smoking: he will less likely to start smoking and more likely to quit smoking with higher education, ceteris paribus. These results are previous literature by Gilman (2008) and Paraje (2017) stating that higher education reduces tobacco consumption.

Even with 12 different independent variables, models 3 can only explain about 20% of the data (according to the R squared, which provide the same information as R square about the portion of explained data). These small pseudo-R squared values indicate that we still need to worry about omitted variable bias since there are a lot of explanatory variables that are not included in the models in this analysis.

The purpose of introducing model 4 is to compare the performance of probit regression and OLS regression on the long regression (models 3). For both smoking and quitting model, the respective coefficients estimated by probit and LMS are very closed, with an exception of *resthome.* However, since the dependent variables are categorical, probit models are preferred to the LMS models to overcome the problem of using a liner to fit a nonlinear relationship, unbounded functional form, and heteroscedasticity that faces by LMS.

Unlike the average marginal effect of continuous variables, the average marginal effect of dummy variables is harder to interpret. Indeed, the effect of dummy variables is meaningful only when the value of that dummy variable is either 0 or 1 since the individual can only have the character associated with the dummy variable or not. Therefore, the average marginal effect does not help us understand the effect of having a particular characteristic on smoking/quitting behavior. To understand the effect of dummy variable correctly, hypothetical cases can be created to predict the change in smoking/quitting probability associating with adopting a particular characteristic. For example, to understand the effect of home restriction on smoking/quitting probability, 2 hypothetical cases are created: case 1 without smoking at

**Table 3.** The result of smoking regression models

```
Regress Smoke on Xs
```

|  | Smoke1 | PPI_1 | Smoke2 | PPI_2 | Smoke3 | PPI_3 | Smoke_OLS |
|---|---|---|---|---|---|---|---|
| main |  |  |  |  |  |  |  |
| educyear | -0.0912*** | -0.0213*** | -0.121*** | -0.0223*** | -0.129*** | -0.0230*** | -0.0228*** |
|  | (0.00229) | (0.00054) | (0.00547) | (0.00103) | (0.00658) | (0.00117) | (0.00119) |
| resthome |  |  | -1.423*** | -0.263*** | -1.370*** | -0.245*** | -0.441*** |
|  |  |  | (0.03571) | (0.00575) | (0.03668) | (0.00589) | (0.01256) |
| restpub |  |  | -0.0415 | -0.00767 | -0.0384 | -0.00687 | -0.00677 |
|  |  |  | (0.03548) | (0.00655) | (0.03665) | (0.00655) | (0.00750) |
| lnprice |  |  |  |  | 0.0392 | 0.00701 | 0.00164 |
|  |  |  |  |  | (0.07393) | (0.01321) | (0.01304) |
| famincome |  |  |  |  | -0.0325*** | -0.00580*** | -0.00568*** |
|  |  |  |  |  | (0.00394) | (0.00070) | (0.00065) |
| age |  |  |  |  | 0.0477*** | 0.00853*** | 0.00862*** |
|  |  |  |  |  | (0.00710) | (0.00126) | (0.00122) |
| agesq |  |  |  |  | -0.000620*** | -0.000111*** | -0.000108*** |
|  |  |  |  |  | (0.00008) | (0.00001) | (0.00001) |
| male |  |  |  |  | 0.101*** | 0.0180*** | 0.0179*** |
|  |  |  |  |  | (0.02857) | (0.00513) | (0.00524) |
| white |  |  |  |  | -0.180* | -0.0321* | -0.0429* |
|  |  |  |  |  | (0.08872) | (0.01588) | (0.02079) |
| black |  |  |  |  | -0.593*** | -0.106*** | -0.120*** |
|  |  |  |  |  | (0.09886) | (0.01762) | (0.02194) |
| asian |  |  |  |  | -0.541*** | -0.0967*** | -0.0939*** |
|  |  |  |  |  | (0.11775) | (0.02101) | (0.02228) |
| hisp |  |  |  |  | -0.606*** | -0.108*** | -0.107*** |
|  |  |  |  |  | (0.05693) | (0.00988) | (0.00834) |
| _cons | 0.199*** |  | 1.845*** |  | 1.524*** |  | 0.825*** |
|  | (0.03125) |  | (0.08152) |  | (0.23034) |  | (0.04472) |
| N | 63489 | 63489 | 19520 | 19520 | 19520 | 19520 | 19520 |
| R-sq |  |  |  |  |  |  | 0.222 |
| F |  |  |  |  |  |  | 230.8 |

```
Standard errors in parentheses
* p<0.05, ** p<0.01, *** p<0.001
```

**Table 4.** The result of quitting regression model.

```
Regress Quit on Xs
```

| | Quit1 | PPI_1 | Quit2 | PPI_2 | Quit3 | PPI_3 | Quit4(OLS) |
|---|---|---|---|---|---|---|---|
| main | | | | | | | |
| educyear | 0.111*** | 0.0417*** | 0.122*** | 0.0406*** | 0.114*** | 0.0353*** | 0.0353*** |
| | (0.00420) | (0.00149) | (0.00932) | (0.00294) | (0.01010) | (0.00302) | (0.00305) |
| resthome | | | 1.073*** | 0.358*** | 1.120*** | 0.347*** | 0.381*** |
| | | | (0.04462) | (0.01233) | (0.04776) | (0.01246) | (0.01458) |
| restpub | | | 0.202*** | 0.0672*** | 0.131** | 0.0407** | 0.0423** |
| | | | (0.04716) | (0.01562) | (0.04907) | (0.01517) | (0.01591) |
| lnprice | | | | | 0.206* | 0.0638* | 0.0599* |
| | | | | | (0.09712) | (0.03008) | (0.02963) |
| famincome | | | | | 0.0247*** | 0.00767*** | 0.00803*** |
| | | | | | (0.00508) | (0.00157) | (0.00158) |
| age | | | | | 0.0257*** | 0.00795*** | 0.00801*** |
| | | | | | (0.00149) | (0.00043) | (0.00044) |
| male | | | | | 0.159*** | 0.0493*** | 0.0487*** |
| | | | | | (0.03815) | (0.01177) | (0.01184) |
| white | | | | | 0.158 | 0.0491 | 0.0571 |
| | | | | | (0.11383) | (0.03524) | (0.03685) |
| black | | | | | -0.0311 | -0.00963 | -0.00540 |
| | | | | | (0.13278) | (0.04114) | (0.04258) |
| asian | | | | | -0.149 | -0.0462 | -0.0322 |
| | | | | | (0.16740) | (0.05187) | (0.05437) |
| hisp | | | | | 0.210** | 0.0652** | 0.0612* |
| | | | | | (0.07593) | (0.02350) | (0.02507) |
| _cons | -1.235*** | | -2.446*** | | -4.220*** | | -0.832*** |
| | (0.05592) | | (0.13445) | | (0.26081) | | (0.07651) |
| N | 25402 | 25402 | 7213 | 7213 | 7213 | 7213 | 7213 |
| R-sq | | | | | | | 0.240 |
| F | | | | | | | 229.9 |

Standard errors in parentheses
* p<0.05, ** p<0.01, *** p<0.001

Home restriction (*resthome* = 0) and case 2 with smoking at home restriction (*resthome* = 1). Other information about these 2 cases, including *education year, smoking restriction in public,* Ln cigarette price, family income, age, gender, race, and ethnicity are identical at take the value at its mean. The predictive smoking/quitting probability of these 2 cases are presented in **table 5**:

**Table 5.** Two hypothetical cases to understand the effect of smoking restriction at home on smoking/quitting probability.

*Smoking*

| _at | Margin | Delta-method Std. Err. | z | P>|z| | [95% Conf. Interval] |
|---|---|---|---|---|---|
| 1 | .4926331 | .0137093 | 35.93 | 0.000 | .4657633  .5195029 |
| 2 | .082486 | .0025668 | 32.14 | 0.000 | .0774552  .0875167 |

*Quitting*

| _at | Margin | Delta-method Std. Err. | z | P>|z| | [95% Conf. Interval] |
|---|---|---|---|---|---|
| 1 | .2638704 | .0137981 | 19.12 | 0.000 | .2368266  .2909142 |
| 2 | .6874929 | .0074796 | 91.92 | 0.000 | .6728332  .7021526 |

**Table 5** indicates that, when other included explanatory variables are at their mean, if there is no smoking restriction at home, the predicted smoking probability is 49.26% (case 1). However, when there is a smoking restriction, the predicted smoking likelihood drops down to 8.25% (case 2). Therefore, the smoking restriction at home reduces 41.01 percentage points, when other included explanatory variables are at the mean. of smoking probability. Similarly, the smoking restriction at home increases the quitting likelihood by 42.36 percentage points, when other included explanatory variables are at their mean. Smoking regulation at home has a significant impact on smoking/quitting decision. This result agrees with what was found by Chapman (1999), Farkas (1999), Gilpin (1999), and Wakefield (2000). However, contrast to their argument that public regulation has stronger power than home regulation in suppressing smoking behavior, this analysis finds that for both smoking and quitting model, the smoking regulation at home has a statistically significant and economic important effect on smoking/quitting decision. In addition, this study suggests that public regulation does not have a statistically significant effect on the smoking decision. The difference between this paper and the previous study in 1999 – 2000 may come from the time difference. It intuitively makes sense that many characteristics associated with a particular specific generation may not be true for the next generation. Further studies are needed to fully address this question.

To further illustrate the effect of education on smoking/quitting decision, the predicted smoking/quitting probabilities at different levels of education are computed and visualized (**Figure 4)**. These predicted values are calculated under 2 conditions: with the smoking restriction at home and without the restriction at home. The down slope of two curves in general in **figure 4a** clearly indicates that education is strongly associated with smoking probabilities. The higher education one has, the less likely he will smoke, holding other included explanatory variables constant. The important role of home regulation is visualized by the gap between two curves. The wider the gap, the more important smoking regulation at home is. The effect of home regulation can vary depending on the level of education of individual since the width of the graph diminishes with the increase of education year. In general, family regulation plays a higher important role in protecting the individual against smoking at a lower level of education. This can be the fact that at a lower level of education associates with lower age, and the younger an individual is, the stronger he is affected by his family.

**Figure 4b** presents the predicted quitting likelihoods by education year with and without smoking regulation at home. The more education an individual has, the more likely he quits smoking. In addition, home regulation promotes the quitting decision. If one does not allow to smoke in his house, he is more likely to quit smoking. The effect of home regulation is presented by the gap between two curves in **figure 4b.** Unlike smoking probabilities, there is not much change in the gap of quitting probabilities with the different family regulation condition, indicating that family regulation has almost similar effect on quitting decision at every level of education, holding other included independent variables constant.

Holding other included independent variables constant, home regulation and education together generate a powerful protective level for an individual from smoking disease: If he has a bachelor degree (16 years of education) and he does not allow to smoke in his house, his predicted probability to start smoking is less than 10% and his predictive likelihood to quit smoking (if he smokes) is more than 70%. If he has a doctoral (about 20 years of education) degree and he does not allow to smoke at home, he is likely to never start to smoke (the predicted probability to start smoking approach 0); however, if he smokes, his likelihood to quit smoking is more than 80%.

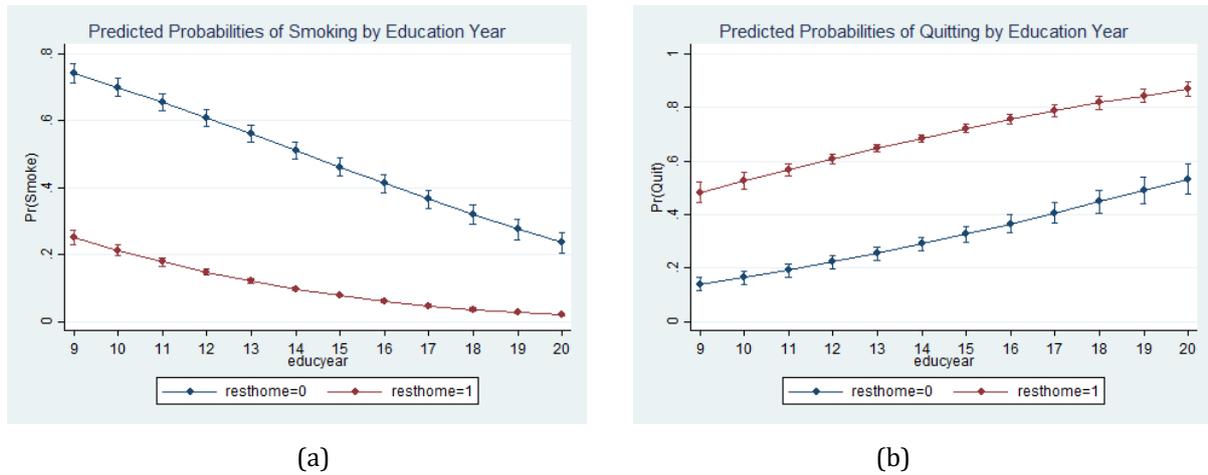

(a)                          (b)

**Figure 4.** Predicted smoking and quitting probabilities by education year. As number of year of education increase in the presence of smoking regulation at home, the smoking probability approach 0 and the quitting probability approach 90%, holding other included independent variables constant.

**IV. Conclusion**

Based on the data from IPUMS CPS 2015, this study shows that education, along with smoking restriction at home, has a strong relationship with smoking/quitting behavior after statistical controls for public restriction, cigarette price, family income, age, gender, race, and ethnicity. Every additional year of education will reduce the 2.3 percentage point of the smoking probability and will add 3.53 percentage point in quitting probability, holding home restriction, public restriction, cigarette price, family income, age, gender, race, and ethnicity constant.

However, the mechanism of education effect on smoking has not been fully developed yet. Many critics argue that since smoking behavior starts early in the lifetime of an individual, at which he did not have a sufficient amount of education, education cannot be the cause of low smoking probability. Maralani (2018) argues that education should be understood as "a bundle of advanced statutes that is developed in childhood." Indeed, higher education correlates to a set of factors that rather than the reason for smoking/quitting behavior. Unfortunately, a clear causal relationship between education and smoking/quitting behavior cannot be established in this analysis, due to the disadvantaged nature of the observational study. In addition, this study suffers from omitted variables bias, even after introducing 11 statistical controls. Indeed, there are many factors that are shown to correlate to education and/or smoking and quitting behavior, such as the smoking status of sibling and parent, the education level of paper, the stress level of the individual, and his social status. However, due to the limitation of data, these variables cannot be incorporated into the model in this analysis, leading to the fact that the estimated coefficients in the model may be biased. This is another disadvantage of an observational study.

The effect of education on smoking/quitting behavior can be explained by the Theory of Rational Addiction that, with higher education, individuals will have a better understanding of detrimental future losses generated by smoking activity and, therefore, will less likely to consume cigarettes. Although there is not a strong evidence to suggest that higher education causes lower smoking probability and higher quitting likelihood, this study offers a relationship between education and smoking/quitting decision that is the foundation for a better controlled study about the effect of education and smoking/quitting decision in the future. Although more research is needed to find out the causal relationship between smoking/quitting and education, this study brings us one step closer to the victory in the battle against the tobacco epidemic.